\documentclass[twocolumn,showpacs,amsmath,amssymb,superscriptaddress,aps,pra]{revtex4-1}
\usepackage[T1]{fontenc}
\usepackage[latin9]{inputenc}
\usepackage{amsmath}
\usepackage{amssymb}
\usepackage{graphicx}
\usepackage{ifthen}
\usepackage{lmodern}

\makeatletter
 
 \@ifundefined{textcolor}{}
 {%
   \definecolor{BLACK}{gray}{0}
   \definecolor{WHITE}{gray}{1}
   \definecolor{RED}{rgb}{1,0,0}
   \definecolor{GREEN}{rgb}{0,1,0}
   \definecolor{BLUE}{rgb}{0,0,1}
   \definecolor{CYAN}{cmyk}{1,0,0,0}
   \definecolor{MAGENTA}{cmyk}{0,1,0,0}
   \definecolor{YELLOW}{cmyk}{0,0,1,0}
 }

\usepackage{dcolumn}
\usepackage{bm}
\usepackage{color}\usepackage{ulem}

\newcommand{\op}[1]{\hat{#1}}
\newcommand{\dagop}[1]{\hat{#1}^{\dagger}}
\newcommand{\bo}[1]{{\mathbf{#1}}}
\newcommand{\mc}[1]{{\mathcal{#1}}}
\newcommand{\wt}[1]{{\widetilde{#1}}}

\makeatother

\begin{document}

\title{Anisotropy in $s$-wave Bose-Einstein condensate collisions and its relationship to superradiance}

\author{P.~Deuar}
\email{deuar@ifpan.edu.pl}
\affiliation{Institute of Physics, Polish Academy of Sciences, Al. Lotnik\'{o}w 32/46, 02-668 Warsaw, Poland.}

\author{J.-C.~Jaskula}
\altaffiliation{Current Address: Harvard-Smithsonian Center for Astrophysics, Cambridge, MA 02138.} 
\affiliation{Laboratoire Charles Fabry de l'Institut d'Optique, CNRS, Universit\`{e} Paris-Sud, Campus Polytechnique RD128 91127, Palaiseau, France.}

\author{M.~Bonneau}
\altaffiliation{Current Address: Vienna Center for Quantum Science and Technology, Atom Institut, TU Wien, Stadionallee 2, 1020 Vienna, Austria.} 
\affiliation{Laboratoire Charles Fabry de l'Institut d'Optique, CNRS, Universit\`{e} Paris-Sud, Campus Polytechnique RD128 91127, Palaiseau, France.}

\author{V.~Krachmalnicoff}
\altaffiliation{Current Address: Institut Langevin, ESPCI ParisTech \& CNRS UMR 7587, 1 rue Jussieu, 75005 Paris, France.} 
\affiliation{Laboratoire Charles Fabry de l'Institut d'Optique, CNRS, Universit\`{e} Paris-Sud, Campus Polytechnique RD128 91127, Palaiseau, France.}

\author{D.~Boiron}
\author{C.~I.~Westbrook}
\affiliation{Laboratoire Charles Fabry de l'Institut d'Optique, CNRS, Universit\`{e} Paris-Sud, Campus Polytechnique RD128 91127, Palaiseau, France.}

\author{K.~V.~Kheruntsyan}
\affiliation{School of Mathematics and Physics, University of Queensland, Brisbane, Queensland 4072, Australia.}

\date{\today}
\begin{abstract}
We report the experimental realization of a single-species atomic four-wave mixing process with BEC collisions for which the angular distribution of scattered atom pairs is not isotropic, 
despite the collisions being in the $s$-wave regime. Theoretical analysis indicates that this anomalous behavior can be explained by the anisotropic nature of the gain  in the 
medium.
There are two competing anisotropic processes: classical trajectory deflections due to the mean-field potential, and Bose
enhanced scattering which bears similarity to super-radiance. We analyse the relative importance of these processes in the
dynamical buildup of the anisotropic density distribution of scattered atoms, and compare to optically pumped super-radiance.
\end{abstract}


\maketitle

\section{Introduction}

Colliding Bose-Einstein condensates (BECs) constitute an atomic four-wave mixing process closely analogous to that in
non-linear optics. 
Of particular interest is the coherent amplification of matter waves \cite{Deng:99,Vogels:02,Perrin:08,RuGway:11},
and the generation of pair correlated atoms \cite{Perrin:07,Ogren-FWM:09,Jaskula:10,Kheruntsyan:12,Vassen:12}.
Atoms scattered during the collision appear in the form of a spherical shell (a ``scattering halo''), with strong correlations in
diametrically opposed regions. 
In the \textit{spontaneous} scattering regime of small halo density, the atom pairs are promising for research into the
fundamentals of quantum mechanics with ensembles of massive particles \cite{Reid:09,Lewis-Swan:13,Lewis-Swan:14}. 
Such states might be used to extend the study of the Einstein-Podolsky-Rosen paradox \cite{Horodecki:09,Reid:09,Kofler:12}, local
realism \cite{Kwiat:95} and Bell inequality tests
to superpositions of different mass distributions.
In the \textit{stimulated}, high halo density regime the atoms have potential applications for precision measurements, and
interferometry \cite{Bouyer:97,Dunningham:02,Campos:03,Gross:10}. 
Quantum correlated pairs can allow one to surpass the limit on the precision of parameter estimation allowed by classical
physics \cite{Giovannetti:04,Pezze:09}. 

In a previous paper, we reported a surprising variation
of the radius of the collision halo with the scattering
angle when elongated condensates collide \cite{Krachmalnicoff:10}.
This feature is counterintuitive when the process is viewed as four wave mixing of matter waves, while a discussion in terms
of particles and their kinetic and potential energies leads to a simple explanation of the observations
\cite{Krachmalnicoff:10,Deuar:11b,Bach:02}.
In the present work, we discuss a different aspect of the same data, the anisotropy in the angular distribution of the number
of scattered atoms in the halo.
This observation is a little surprising when viewed as the scattering of particles because the collisions are well into the
$s$-wave regime, where scattering amplitudes are isotropic. 
This regime is to be contrasted with that of higher speed collisions \cite{Thomas:04,Buggle:04}
in which contributions of $d$-wave amplitudes can produce an angular dependence. 
Anisotropies are less surprising in a wave context.
The process of optically pumped superradiance \cite{Inouye:99,Moore:99,Vardi:02,Schneble:03, Sadler:07, Li:08, Hilliard:08,
Buchmann:10,Lopes:14} which occurs when an elongated atomic cloud is illuminated with light, exhibits strong anisotropies
which can be explained in terms of an anisotropic gain medium. 

In this paper we will show that the anisotropy in the scattered atom number can be explained by appealing to an anisotropic
gain, but that the contributing factors differ significantly from those in optically pumped superradiance.
We identify two competing processes: simple classical deflections of particle trajectories and gain enhanced scattering. 
In collisions of highly elongated BECs, the end-fire modes are less dominant because of different energy-momentum conservation relations
compared to atom photon collisions (superradiance) or  molecular dissociation \cite{Vardi:02,Ogren-directionality}.

The structure of the paper is as follows:  we first describe the experiment and the observed anisotropies in Sec.~\ref{EXP}. 
In Sections~\ref{CLASS} and~\ref{PDC}, respectively, we describe the scattering process from two simplified viewpoints: 
that of classical particles rolling on a potential hill formed by the remaining condensate; and that of quantum parametric
down-conversion of bosons. 
These processes are shown to give rise to two competing anisotropies of the scattering halo. 
In Sec.~\ref{STAB} the situation is described using the much more accurate numerical stochastic Bogoliubov treatment which
incorporates both simplified models as special cases. 
This allows us to quantify how the competition between the anisotropies is resolved. 
In Sec.~\ref{SRAD} we compare the system to optically pumped superradiance, and then make concluding remarks in Sec.~\ref{CONC}.


\section{Experiment}
\label{EXP}

\begin{figure}
\includegraphics[width=7cm]{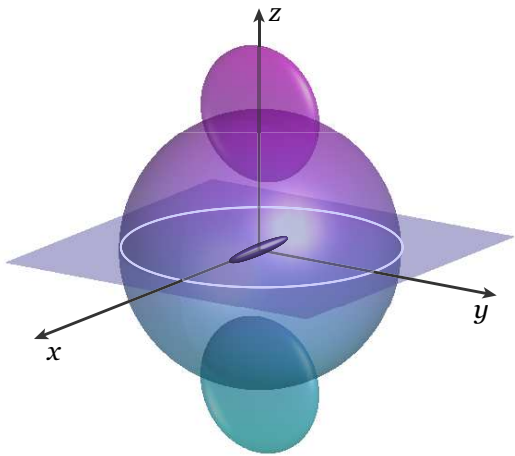} \caption{ (Color online) Schematic diagram of the collision geometry in
the center-of-mass frame in which we denote the collision axis as $z$. 
The two disks represent the colliding condensates after their mean-field induced expansion. 
The sphere represents the halo of scattered atoms. 
The initial, cigar-shaped condensate, whose long axis coincides with $x$, is shown in the center. We analyze the experimental data in the $x-y$ plane. 
}
\label{fig:Axes} 
\end{figure}

The experimental apparatus was described in Ref.~\cite{Krachmalnicoff:10}.
Briefly, we start from a BEC of $\sim10^{5}$ $^4$He atoms magnetically trapped
in the $m_{x}=1$ sublevel of the $2^{3}S_{1}$ metastable state. 
The quantization axis is defined by a bias magnetic field of $\sim0.25$~G along the axial direction of the condensate, labeled $x$ (see Fig.~\ref{fig:Axes}).
The trap is cylindrically symmetric, with axial and radial frequencies of $\omega_{x}/2\pi=47$~Hz and $\omega_{\perp}/2\pi=1150$~Hz, respectively. 
To generate the two colliding BECs, we use a combination of Raman
and Bragg laser pulses \cite{Krachmalnicoff:10} to transfer half of the atoms to a state moving 
at relative velocity of $2v_{0}$ (with respect to the other half) along the $z$-axis. 
In the center-of-mass frame, the 
colliding BECs move at velocities $\pm v_0$, with $v_{0}=7.31$~cm/s (momentum $k_0=mv_0/\hbar=4.61\times 10^6$ m$^{-1}$, in wave-number units),
which is $\sim4$ times the speed of sound in the center of the BEC.
The internal state of the atoms after the transfer is $m_{x}=0$ and therefore the 
atoms become insensitive to the magnetic trapping field. 
The freely colliding condensates thus separate along the $z$ axis and create a
scattering halo that contains about 2000 atoms, 2\% of the initial condensate, and lies in the crossover between the low-occupation spontaneous regime and the stimulated one. 

\begin{figure}[tbp]
\includegraphics[width=0.9\columnwidth,clip]{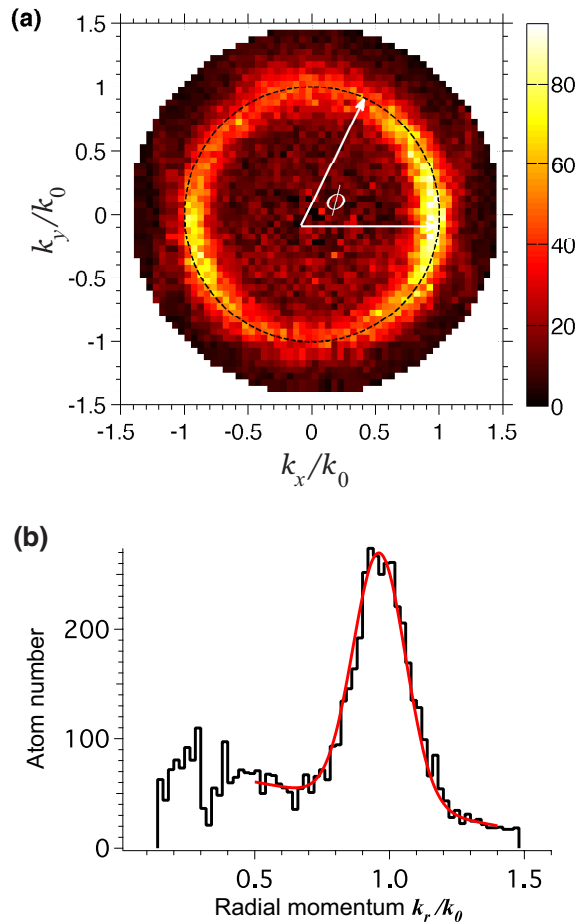} 
\caption{(Color online) 
Experimental data (averaged over $\sim1600$ experimental runs). 
(a) The momentum space density $n(k_{x},k_{y})$ of the
observed scattering halo on the equatorial $k_{x}-k_{y}$ plane (in arbitrary units).
It has been averaged over a disk of thickness $[-0.1k_{0},+0.1k_{0}]$
along $k_{z}$, and the momentum is normalized to the collision momentum $k_{0}=mv_0/\hbar$. 
(b) Atom distribution in the halo as a function of the radial momentum
$k_{r}$. 
We show the sum of counts over all runs for a single azimuthal sector of width $22.5^{\circ}$.
The red line is the fit to Eq.~(\ref{eq:fit}). 
}
\label{fig:DensityData} 
\end{figure}

After the collision (the bulk of which takes $\sim70$ $\mu$s), the atomic cloud expands and the atoms fall onto a microchannel plate detector placed $46.5$ cm below the trap center. 
A delay line anode permits reconstruction of a 3D image of the cloud of atoms in position space. 
The flight time to the detector ($\sim 300$ ms) is long enough that, to a good approximation,
the 3D reconstruction can be traced back to a 3D image of the momentum distribution immediately after the collision, when the mean field energy of the condensate has been converted to kinetic energy.
The collision geometry allows detection of the
halo on the entire plane containing the anisotropy of the BEC (the $x-y$ plane in Fig.~\ref{fig:Axes})
while the condensates themselves are detected well away from the plane. 
Thus, local saturation of the detector by the BECs does not interfere with the analysis of the halo in the $x-y$ plane. 
As in Ref.~\cite{Perrin:07}, we observe a strong correlation between atoms with opposite momenta
confirming that the observed halo is indeed the result of binary collisions.

\begin{figure}[tbp]
\includegraphics[width=\columnwidth,clip]{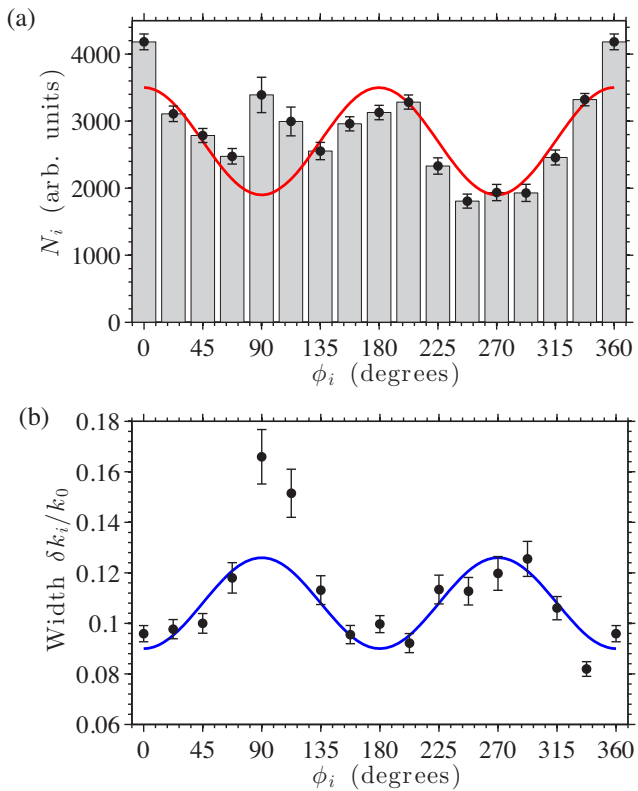}
\caption{ (Color online) 
Halo anisotropy in the experiment. 
Panel (a): the angular distribution of the number of scattered
atoms $N_i$ on the equatorial $k_x-k_y$ plane (calculated from fits to Eq. (\ref{eq:fit}) after counting atoms over all runs).
The solid red curve shows a sinusoidal fit to $N_i$, which is anisotropic and maximal along the direction of the long axis of the condensate at $0^{\circ}$ and $180^{\circ}$.
Panel (b): the fitted radial width $\delta k_i$ of the halo density as a function of the azimuthal angle $\phi_i$. The blue (solid) curve shows a sinusoidal fit. 
In both panels, the angular bins (or azimuthal sectors), labeled by $i=1,2,...16$, are centered at $0^{\circ},22.5^{\circ},45^{\circ},...337.5^{\circ}$. 
}
\label{fig:AtomNbMod} 
\end{figure}

In Fig.~\ref{fig:DensityData}(a) we show a momentum-space slice of the scattering
halo in the $k_x-k_y$ plane that reveals its annular structure. 
The ring shown in Fig.~\ref{fig:DensityData}(a) can be divided into azimuthal sectors 
(sixteen here, labeled $i$) and fitted radially with a Gaussian peak plus a linearly sloped background 
\begin{equation}
n_{i}(k_r)=\alpha_{i}+\beta_{i}(k_r-K_{i})+
\mathcal{A}_{i}\exp\left(-{\frac{(k_r-K_{i})^{2}}{2\:\delta k_{i}^{2}}}\right)
\label{eq:fit}
\end{equation}
as in Fig.~\ref{fig:DensityData}(b) 
to extract the values for the peak local density $\mathcal{A}_{i}$,
the peak radius $K_{i}$, and the radial rms width $\delta k_{i}$ of the halo as functions of the angle $\phi_i$. 
We also obtain the total scattered atom number in sector $i$ by integrating the Gaussian
in Eq. \eqref{eq:fit}. In the limit $\delta k_i\ll K_i$ that applies here, $N_{i}=\sqrt{2\pi}\mathcal{A}_{i}\delta k_{i}$.
We will focus on $\delta k_{i}$ and $N_{i}$ (with the third parameter $\mathcal{A}_{i}$ following from these two in this approximation), while 
analysis of the anisotropy of the radius $K_i$ was reported in an earlier paper \cite{Krachmalnicoff:10}. 
We plot the fitted halo width and the scattered atom number in Fig.~\ref{fig:AtomNbMod} as a function of $\phi$, the angle from the $k_x$ axis.

Both quantities show an anisotropy of a few tens of percent of their mean values. 
Oscillations in the two quantities have the opposite phase, meaning that the anisotropy of the 
peak density $\mathcal{A}_{i}$ is even stronger than for $N_i$, since it is 
proportional to the ratio of the two.
Both curves seem to have two outliers centered at $90^\circ$ and at $112.5^\circ$. As can be seen in Fig.~\ref{fig:DensityData}(a), the halo is very broad at these angles while the density remains significant. We believe that this anomalous behavior is an experimental problem---possibly due to the detector, and will disregard it in this paper. Indeed, the model Hamiltonian and the geometry of the collision have certain reflectional symmetries (e.g., with respect to the $(x-z)$ and $(y-z)$ planes) which preclude any differences between the scattering outcomes at $90^\circ$ and $270^\circ$. This implies that the halo density should not behave otherwise than with an 180 degree periodicity, and we will therefore impose this symmetry in our analysis.

The collision energy in this experiment is low enough that the scattering is well into the $s$-wave regime. 
Therefore, one would naively expect isotropic distributions. Somewhat less naively, an anisotropic momentum distribution of the source clouds such as we have here can cause variations in the width $\delta k$ and peak density $\mathcal{A}$ of the halo that depend on the scattering angle. 
However, without further effects, one would expect the width and peak density variations to balance to produce an isotropic distribution of the atom number $N_i$ in the $s$-wave regime.  
Nevertheless, this is not what is seen. The distribution of the number of scattered atoms in Fig.~\ref{fig:AtomNbMod}(a) is clearly anisotropic.

In the centre-of mass frame, the scattered atom number is higher in the direction of the initial cigar, or in the parlance of superradiance, the direction of the ``end-fire'' modes \cite{Inouye:99}.  
One is therefore immediately tempted to interpret the anisotropy in the number as the result of Bose-enhanced scattering, that is to an anisotropic gain. 
However, before making such an assertion, we must consider whether classical particle-like effects can account for the observed angular distribution.


\section{Anisotropy due to classical-particle effects} 
\label{CLASS}

The scattered atoms are sensitive to the potential created by the mean field of the condensates. 
In \cite{Krachmalnicoff:10}, an anisotropy of the mean halo radius caused by this was analyzed. 
The shift of the radius from $k_0$ was due to a combination of two factors: extra energy needed to scatter into a non-condensed mode, and the partial recovery of this energy when a particle rolls back down the 
mean-field potential created by the remaining condensate \cite{Krachmalnicoff:10,Deuar:11b}. Since this potential is non-isotropic in the present case, we not only expect some anisotropy in the radius, but also the direction of the particles' trajectories as they roll down the mean field potential.  Examples are shown in Fig.~\ref{fig:trajectories} in the Appendix. This may lead to an anisotropy of the number of particles $N_i$, not just their radial position. Naively, one might expect that because of the defocusing nature of the potential, atoms will be pushed away from the condensate axis, an effect that we will call ``reverse anisotropy''.  
This is a deviation from the isotropic case in the reverse sense to that seen in the experiment, and corresponds to
$\Gamma_N<1$ rather than $\Gamma_N>1$ (see Eq. (\ref{GammaN}) below for the definition of $\Gamma_N$).

Predicting the quantitative effect of the mean field on the anisotropies is difficult to do from first principles.
One must include the fact that the atomic density profile changes with time as the two condensates separate and expand, and the fact that the 
initial position and relative velocities of the atoms are not fixed but must be averaged over appropriate (anisotropic) distributions. 
Hence, to assess the effect of the interaction between an atom which has undergone a collision and those left over in the condensates, we have performed a simulation via a classical test-particle method.
It is described in the Appendix, and takes into account the above factors.  

\begin{figure}[tbp]
\includegraphics[width=0.85\columnwidth,clip]{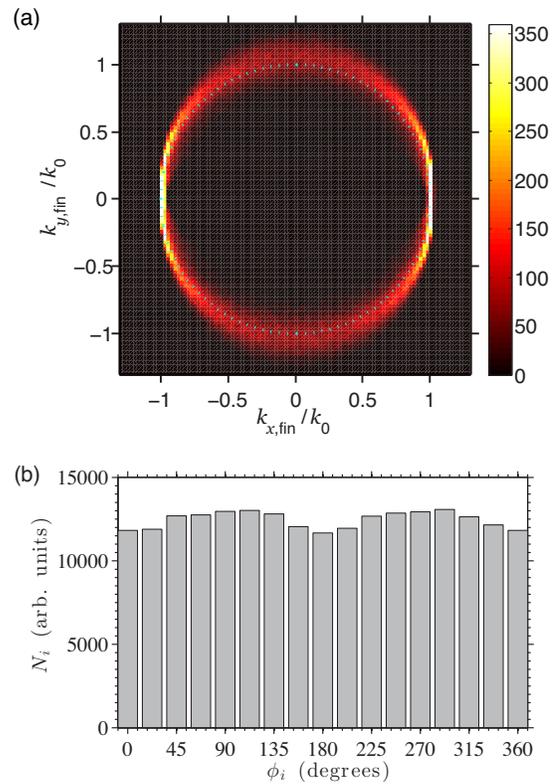}
\caption{(Color online) 
Halo properties in a classical test-particle model: 
(a) A scatter plot of the distribution of final momenta of classical test-particles after their escape from the collision zone. 
(b) The angular distribution of the number of test-particles $N_i$ in azimuthal sectors centered at angles $\phi_i$ ($i=1,2,... 16$).
The key observations are that the atom number depends weakly on the angle, but in the opposite sense to that observed in the experiment, and that the halo width exhibits a strong anisotropy that reflects the shape of the source in momentum space.}
\label{fig:classical} 
\end{figure}

The plot in Fig.~\ref{fig:classical} shows that a careful simulation of the mean field effect on the classical trajectories results in a small anisotropy (of the order of $10$\%, with $\Gamma_N\approx 0.9$) in the number of scattered atoms. 
The classical test-particle method can indeed account for several qualitative features of the experimentally observed collision
halo, such as the density, width, and the mean radius of the scattering shell. 

However, the particle number anisotropy in Fig.~\ref{fig:classical}(b) has the opposite sense in comparison with the observations: the classical simulation predicts slightly more atoms at $90^\circ$ and $270^\circ$, which is the ``reverse  anisotropy'' that we foreshadowed above. 
The Appendix presents details of the calculation, a demonstration of the importance of including the full time-evolution of the source cloud (compare Figures~\ref{fig:classical} and~\ref{fig:classical-static}), as well as some example particle trajectories. 
Also, the classical calculation shows that mean field effects lead to a much stronger anisotropy in the halo width than is observed in the experimental data of Fig.~\ref{fig:AtomNbMod}. 
In any case, something more is needed to explain the experimental observations.

In the remainder of the paper we will discuss the inclusion of quantum effects in the calculation of the scattered atom distribution.


\section{Simple quantum treatment: Relation to parametric down-conversion}
\label{PDC}

We begin by neglecting mean field effects and explore Bose-enhancement and other quantum effects with a simple model inspired by the theory of parametric down-conversion.   

The simplest quantum treatment, in which the high-density regions
in the final momentum distribution as well as the location of maxima
in the binned atom number are predicted to be at $0^{\circ}$ and
$180^{\circ}$, can be accomplished by drawing an analogy of the condensate
collision process with parametric down-conversion from quantum optics.
In this model, the peaks in the binned atom number at $0^{\circ}$
and $180^{\circ}$ would be attributed to parametric amplification
of quantum noise, which can lead to an exponentially growing population
of the ``down converted'' modes due to the effect of Bose stimulation.

For highly anisotropic condensates, as is the case in our collision
experiments, with $0^{\circ}$ and $180^{\circ}$ corresponding to
the long, axial direction, the larger population of the scattering
modes at these angles would be similar to superradiant amplification
of the ``end-fire'' modes in light scattering from an elongated
condensate~\cite{Inouye:99,Schneble:03,Lopes:14}.

The formal analogy of the condensate collisions with multimode parametric
down conversion in the undepleted ``pump'' approximation was discussed
previously in Refs. \cite{Perrin:08,Ogren-FWM:09}. The undepleted ``pump''
refers to the source condensate,  which will be assumed to stay constant
in time. The approximation is applicable to relatively short collision
durations resulting in the conversion of less than 5\% of the atoms
from the source condensate into the scattering modes. In this model,
the scattered atoms can be described by a small fluctuating component
($\widehat{\delta}$) of the field operator $\widehat{\Psi}$, which
is decomposed as in the Bogoliubov approach \cite{Jaskula:10}, $\widehat{\Psi}=\Psi_{0}+\widehat{\delta}$,
where $\Psi_{0}$ is the mean-field component. The corresponding Heisenberg
equations for $\widehat{\delta}$ can be written down as \cite{Ogren-FWM:09}
\begin{equation}
\frac{\partial\widehat{\delta}(\mathbf{x},t)}{\partial t}=i\left[\frac{\hbar\nabla^{2}}{2m}+\frac{\hbar k_{0}^{2}}{2m}\right]\widehat{\delta}(\mathbf{x},t)-i{\cal {G}}(\mathbf{x})\widehat{\delta}^{\dag}(\mathbf{x},t),\label{x-space-eq}
\end{equation}
where $\hbar k_{0}$ is the collision momentum of each condensate,
and ${\cal {G}}(\mathbf{x})=g\rho_0(\mathbf{x})/\hbar$ is an
effective parametric coupling  spatially dependent on the initial condensate density $\rho_0(\bo{x})=|\Psi_0(\mathbf{x},t=0)|^2$. In addition, $g=4\pi\hbar^2a/m$ is the coupling constant describing the $s$-wave scattering interactions, with $a$ being the corresponding scattering length and $m$ -- the atomic mass.

With a further simplifying assumption
of a homogeneous source condensate with density $\rho$, and hence a
spatially-uniform coupling constant ${\cal {G}}$, Eq.~(\ref{x-space-eq})
can be easily solved analytically in momentum space. The solutions
for the occupancies of the plane-wave momentum modes $\mathbf{k}$ have the
following familiar form \cite{Savage:06,Perrin:08,Ogren-FWM:09}:
\begin{equation}
n_{\mathbf{k}}(t)=\frac{{\cal {G}}^{2}}{{\cal {G}}^{2}-\Delta_{k}^{2}}\sinh^{2}\left(\sqrt{{\cal {G}}^{2}-\Delta_{k}^{2}}\ t\right).\label{n-k-analytic-b}
\end{equation}
where $\Delta_{k}$ corresponds to the effective phase mismatch or
the energy offset from the resonance condition, and is given by
\begin{equation}
\hbar\Delta_{k}\equiv\frac{\hbar^{2}|\mathbf{k}|^{2}}{2m}-\frac{\hbar^{2}k_{0}^{2}}{2m}.\label{eq:resonance}
\end{equation}
From Eq.~(\ref{n-k-analytic-b}) we see that the modes with ${\cal {G}}^{2}-\Delta_{k}^{2}>0$
can experience Bose enhancement and grow exponentially with time,
whereas the modes with ${\cal {G}}^{2}-\Delta_{k}^{2}<0$ oscillate
at the spontaneous noise level. The absolute momenta of the exponentially
growing modes lie near the resonant momentum $\hbar k_{0}$, and therefore
the condition ${\cal {G}}^{2}-\Delta_{k}^{2}>0$ can be used to define
an approximate width of the $s$-wave scattering shell \cite{Perrin:08,Bach:02}.
Such a definition gives a power-broadened width and explains why the
experimentally observed width along the $x$ axis is larger than that seen with classical test-particles (Fig.~\ref{fig:classical}), which followed directly from the width of the source momentum distribution.

The same Eq.~(\ref{n-k-analytic-b}) can be used to estimate the
occupation numbers of the resonant modes propagating along $x$ (i.e.
at $0^{\circ}$ and $180^{\circ}$ angles) and along the transverse
direction $y$. In such an estimate, one assumes that the above results
can be applied to a finite-size, box like system whose length along
$x$ is 
larger than along $y$. In this case, Eq.~(\ref{n-k-analytic-b})
should be applied for as long as the scattering modes $\mathbf{k}$
``see'' and propagate within the parametric gain medium characterized
by the coupling constant ${\cal {G}}$. In the quantum optical analog,
the corresponding time-scale is defined by the geometric size of the
nonlinear crystal and hence a finite propagation time of a light pulse
within the crystal. In the present case, the role of the gain medium
is taken by the source condensate, and for 
anisotropic condensates
the propagation times along $x$ and $y$ can be quite different.
Accordingly, the occupation numbers of the modes propagating along
the long axis $x$ can be 
significantly larger than those propagating along
$y$, explaining qualitatively why the experimentally observed atom
numbers in annular bins at $0^{\circ}$ and $180^{\circ}$ are higher
than at $90^{\circ}$ and $270^{\circ}$.

For more quantitative estimates, one should take into account the
fact that the role of the gain medium for the scattering modes propagating
on the $x-y$ plane is taken not by the initial source condensate itself,
but by the overlap region between the two colliding condensates. In
this picture, the anisotropy of the gain medium is defined by the
shape of this overlap region and the actual escape time of the scattered
atoms from the overlap zone. For estimating the escape time $t_{x}$
and $t_{y}$ along the $x$ and $y$ directions, we use a simple mean-field
model in which the condensate density profiles are approximated by
an inverted Thomas-Fermi parabola, while the spatial separation is
taken into account via the center-of-mass dynamics of counter-propagating
clouds at momenta $\pm k_{0}$ along the $z$ direction undergoing a simultaneous self-similar expansion
of the condensates as in Ref.~\cite{Castin-Dum}.

We estimate that the resonant modes propagating
along $y$ with momentum $k_{0}$ escape the overlap
zone on a time-scale of $t_{y}=40$ $\mu$s in our system, while the escape time
$t_{x}$ along the longitudinal axis $x$ is determined (for
our geometry) simply by the time required for complete spatial separation
of the two condensates. For our parameters it is equal to $t_{x}\approx70$
$\mu$s. As an approximate estimate of ${\cal{G}}$ we use the peak
density of the initial source condensate $\rho_0(0)=2.4\times10^{19}$
atoms/m$^{3}$, giving ${\cal{G}}\approx3.6\times10^{4}$ s$^{-1}$.
In order to account for the fact that the actual density of the inhomogeneous
condensate becomes smaller as one moves away from the center, we apply
Eq.~(\ref{n-k-analytic-b}) --- only as a crude estimate --- to half of
the durations $t_{x}$ and $t_{y}$. Accordingly, we obtain ${\cal{G}}t_{y}/2\approx0.72$
and ${\cal{G}}t_{x}/2\approx1.26$ and therefore the respective maximum mode populations
can be estimated as $n_{k_{0,y}}(t_{y}/2)=\sinh^{2}({\cal{G}}t_{y}/2)\approx0.61$
and $n_{k_{0,x}}(t_{x}/2)=\sinh^{2}({\cal{G}}t_{x}/2)\approx2.6$. 

These estimates agree with the anisotropic trend seen in the experiment. They
should still be regarded as very crude approximations since the
actual density in the condensates overlap zone varies not only spatially,
but also with time, and goes down to zero on a timescale of $70$ $\mu$s. 
The resulting dependence of the effective parametric
gain ${\cal {G}}(\textbf{x},t)$ leads to mode mixing and makes the simple analytic
solution (\ref{n-k-analytic-b}) overestimate
the actual mode occupation numbers.
We can conclude that the anisotropic parametric amplification plays a role in generating the observed anisotropy in the halo density, but 
does not fully explain the quantitative aspects. 

The picture
that emerges from our analysis so far is that the classical trajectory
deflections and parametric amplification are competing processes,
and the full understanding of the collision halo requires a more advanced
quantum treatment that incorporates both processes. 
 One might expect that the parametric amplification becomes more dominant over classical effects as cloud density grows. 


\section{Detailed quantum treatment: stochastic Bogoliubov}
\label{STAB}

We now turn to
the analysis of the collision dynamics within the much more quantitatively accurate stochastic Bogoliubov
method. It includes a variety of processes including both competing ones mentioned in the previous section, as well as incorporating the temporal evolution of the condensate mean-fields. It allows us to study the competition between deflections and parametric amplification and to observe how the halo anisotropy forms.

The method  is described in detail in Ref. \cite{Deuar:11}, and was used previously for simulating our experiments and for other studies \cite{Krachmalnicoff:10,Jaskula:10,Kheruntsyan:12,Deuar:13,Lewis-Swan:14}.
The approach boils down to evolving the system in a time-dependent Bogoliubov approximation, taking the condensate part $\Psi_0(\bo{x},t)$ at time $t$ as the solution of the full Gross-Pitaevskii mean-field evolution equation for the colliding condensates:
\begin{equation}\label{GPeq}
  i\hbar\,\frac{\partial\Psi_0(\bo{x},t)}{\partial t}= \left[-\frac{\hbar^2}{2m}\nabla^2 + g|\Psi_0(\bo{x},t)|^2\right]\Psi_0(\bo{x},t).
\end{equation}
The scattered atoms are described using a bosonic field operator $\hat\delta(\bo{x,t})$, and following the Bogoliubov approach we use the linearized equations of motion for this field. At the same time, we assume that the proportion of scattered atoms is small, so that the self-interaction of $\op{\delta}$ can be neglected, leading to
\begin{eqnarray} \label{Bog}
  i \hbar\,\frac{\partial\op{\delta}(\bo{x},t)}{\partial t}&=&\left[ -\frac{\hbar^2}{2m}\nabla^2+ 2g|\Psi_0(\bo{x},t)|^2\right] \op{\delta}(\bo{x},t)\nonumber\\
  &&+g\Psi_0(\bo{x},t)^2 \op{\delta}^\dagger(\bo{x},t).
\end{eqnarray}
In comparison with the undepleted pump model (\ref{x-space-eq}),  this implements a full time-dependent Bogoliubov description. The differences have been analyzed e.g. in \cite{Deuar:11,Deuar:13}.
The numerical lattice required to describe this model is too large for a direct solution of the Bogoliubov-de Gennes equations corresponding to (\ref{Bog}) to be tractable, so in the stochastic Bogoliubov approach, $\op{\delta}$ is represented using the positive-$P$ representation. In this, we sample the distribution of two complex fields $\delta(\bo{x},t)$ and $\wt{\delta}(\bo{x},t)$ that represent $\op{\delta}$ and $\dagop{\delta}{}$ and obey the following linear It\={o} stochastic differential equations that can be numerically integrated \cite{Krachmalnicoff:10,Deuar:11}:
\begin{subequations}\label{STABeq}
  \begin{eqnarray}
    i\hbar\,\frac{\partial\delta({\bo{x}},t)}{\partial t}&=&\left(-\frac{\hbar^2}{2m} \nabla^2 + 2g|\Psi_0({\bo{x}},t)|^2\right)\delta({\bo{x}},t)\nonumber\\
    &&\hspace*{-0.9cm}+g\,\Psi_0({\bo{x}},t)^2\,\wt{\delta}({\bo{x}},t)^*+\sqrt{i\hbar g}\,\Psi_0({\bo{x}},t)\,\xi({\bo{x}},t),\qquad\label{STABpsib}\\
    i\hbar\,\frac{\partial\wt{\delta}({\bo{x}},t)}{\partial t}&=&\left(-\frac{\hbar^2}{2m} \nabla^2 + 2g|\Psi_0({\bo{x}},t)|^2\right)\wt{\delta}({\bo{x}},t)\nonumber\\
    &&\hspace*{-0.9cm}+g\,\Psi_0({\bo{x}},t)^2\,\delta({\bo{x}},t)^*+\sqrt{i\hbar g}\,\Psi_0({\bo{x}},t)\,\wt{\xi}({\bo{x}},t).\qquad\label{STABpsitb}
  \end{eqnarray}
\end{subequations}
Here $\xi({\bo{x}},t)$ and $\wt{\xi}({\bo{x}},t)$ are independent, real stochastic Gaussian noise fields with zero mean and 
\begin{equation}
  \langle \xi({\bo{x}},t)\xi({\bo{x}}',t') \rangle
  = \langle \wt{\xi}({\bo{x}},t)\wt{\xi}({\bo{x}}',t') \rangle
  =\delta^{(3)}({\bo{x}}-{\bo{x}}')\delta(t-t')
\end{equation}
are the only nonzero second moments. 
The ensemble of such stochastic realizations corresponds to the full Bogoliubov dynamics (\ref{Bog}), and allows one to estimate observables to within a well defined statistical accuracy. 
Any observable that can be expressed in terms of normally-ordered operator products can be calculated by an appropriate stochastic average over the ensemble: 
$\langle(\dagop{\delta})^n(\op{\delta})^m\rangle = \left\langle\textrm{Re}\left[(\wt{\delta}^{\ast})^n\delta^m\right]\right\rangle_{\rm stoch}\ $. 

The initial state is a superposition of two counter-propagating, mutually coherent atomic clouds created at $t=0$:  
$\Psi_0(\bo{x},t=0) = \sqrt{\rho_0(\bo{x})/2}\,[e^{ik_0 z}+e^{-ik_0 z}]$, 
and a vacuum in the scattered field $\op{\delta}(\bo{x})$. $\rho_0(\bo{x})$ is the GP ground state density in the trap at $t=0$. 
We simulated the collision out to a time of 240 $\mu$s, for a variety of total atom numbers $N$.

\subsection{Comparison to experiment}

\begin{figure}
\includegraphics[width=\columnwidth]{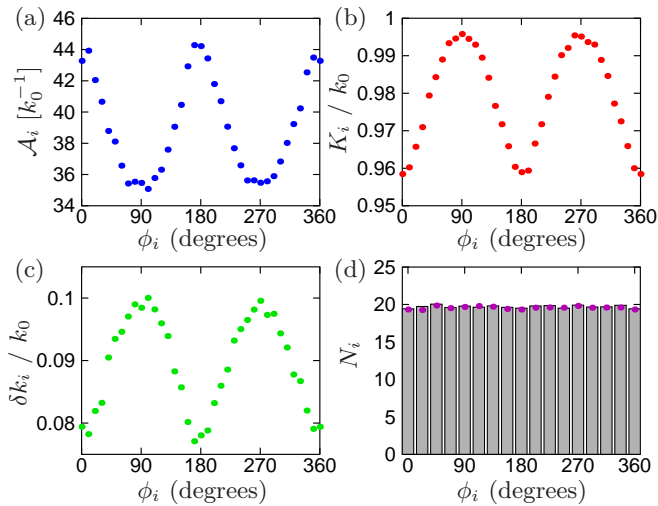}
\caption{(Color online) Direction-dependent properties of the halo from the stochastic Bogoliubov simulation on the $k_x-k_y$ plane after $96$~$\mu$s for $N=10^5$ atoms. The 
local peak density $\mc{A}_i$ (a), mean halo radius $K_i$ (b), and thickness $\delta k_i$ (c) come from a fit of the linear density $n_i(k_r)$ in each sector (of width $10^{\circ}$) to Eq. (\ref{eq:fit}) with no background ($\beta_i=\alpha_i=0$). The sectors divide up a disk around the plane in the interval $k_z\in[-0.2k_0, +0.2k_0]$. Panel (d) shows the number of scattered atoms $N_i$ ($i=1,2,...16$) per azimuthal sector centered at $\phi_i$ from a direct count (histogram), and from the simple estimate $\sqrt{2\pi}\mc{A}_i\delta k_i$ (purple circles). Statistical fluctuations are visible.
}
\label{fig:n1e5bog} 
\end{figure}

Taking the typical value of $N=10^5$ atoms, the angular modulation of the halo in the $k_x-k_y$ plane after 96 $\mu$s is shown in Fig.~\ref{fig:n1e5bog}. To reduce statistical fluctuations, the density is averaged over the range $k_z\in[-0.2k_0, +0.2k_0]$. 
Similarly to the experiment, the local peak density, radius and thickness come from a fit to (\ref{eq:fit}) but without the sloped or constant backgrounds ($\beta_i=\alpha_i=0$) since technical noise is absent here.  

The amplitude $\mathcal{A}_i$ peaks at $0^{\circ}$ and $180^{\circ}$, while the radius and width are narrowed at these angles. This matches the angle-dependent variations seen in the experimental data of Fig.~\ref{fig:DensityData}(a) and \ref{fig:AtomNbMod}(b).
The calculated and observed widths are similar except for the anomaly in the experimental data around $\phi=90^{\circ}$. 
We can also compare the absolute scattered atom number between the simulation and the experiment. Taking into account the 12\% detection efficiency \cite{Jaskula:10} and compensating for different sector widths and $k_z$ averaging ranges, the average value of $N_i\approx8.7$ per trial seen in Fig.~\ref{fig:n1e5bog}(d) corresponds to an expected 
1900 atom counts in the 1600 experimental trials.  This is in agreement  with Fig.~\ref{fig:AtomNbMod}(a) in the $90^{\circ}$ and $270^{\circ}$ directions, but the simulation does not reproduce the marked variation in atom number shown in Fig.~\ref{fig:AtomNbMod}(a).

\subsection{Deflection versus amplification}

\begin{figure}
\includegraphics[width=0.98\columnwidth]{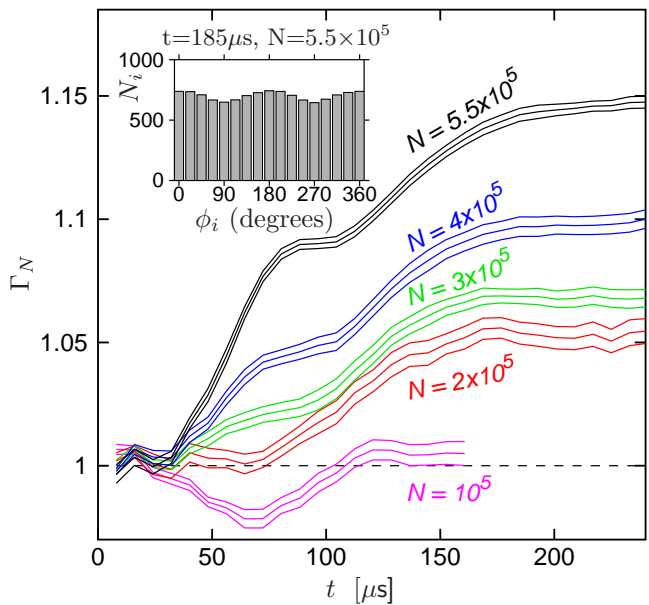}
\caption{(Color online) The relative anisotropy of the scattered atom number $\Gamma_N$ as a function of time $t$ from the stochastic Bogoliubov simulations. $N$ is the total atom number in the main cloud. Triple lines show the statistical uncertainty. 
The inset shows the angular variation of the scattered atom count $N_i$ ($i=1,2,...16$) per azimuthal sector centered at $\phi_i$ for the highest density case 
at the end of the collision, as per Fig.~\ref{fig:n1e5bog}(d). 
}
\label{fig:Ndiffbog} 
\end{figure}

In order to study the atom number anisotropy further, we can vary the parameters in the simulation. We 
define a relative anisotropy $\Gamma_N$ of the number of scattered atoms per sector $N_i$
in terms of the ratio between the values in the $k_x$ direction (peak in the experiment) and those along the $k_y$ direction (minimum in the experiment): 
\begin{equation}\label{GammaN}
\Gamma_N = \frac{N_i(\phi=0^{\circ},180^{\circ})}{N_i(\phi=90^{\circ},270^{\circ})}
\end{equation}
To reduce statistical noise, we average the values obtained for three of the $10^{\circ}$ bins nearest to the axes to estimate the peak and minimum values.  
A direct count of the number of scattered atoms is used. 
Fig.~\ref{fig:Ndiffbog} shows the calculated  anisotropy as a function of time and atom number. This reveals the competition between classical trajectory deflections and parametric amplification because 
the strength of the two effects scales differently with time and $N$.

The most visible feature on this diagram is that the long-time anisotropy grows with increasing total atom number in the source condensate. This is indicative of the parametric amplification of Sec.~\ref{PDC}. A larger total atom number for the same trap frequencies implies a higher peak density, therefore a larger gain coefficient $\cal{G}$, leading finally to greater anisotropy via Eq.~(\ref{n-k-analytic-b}).

Several other lines of evidence confirm the presence of Bose-enhanced scattering: 
The markedly non-sinusoidal dependence seen in Figs.~\ref{fig:n1e5bog}(a-c) and \ref{fig:Ndiffbog}(inset) is a characteristic indicator of the presence of the parametric amplification process. The geometry of the collision is such that under the same assumptions as in Sec.~\ref{PDC}, 
a particle scattered at $t\approx0$ with zero momentum in the $z$ direction in the center-of-mass frame has a travel time 
$t_{\phi} \approx t_y \sqrt{1+\cos^2\phi}$ 
for clouds that are strongly elongated along $x$. Hence, the angular dependence of the peak halo density will vary as 
\begin{equation}\label{Aphi}
\mc{A}(\phi) \sim \sinh^2\left[ C \sqrt{3+\cos2\phi}\,\right].
\end{equation}
where $C\propto\mc{G}$, and the modulation will become visibly non-sinusoidal when $C\gtrsim1$. 

We have also made calculations using the perturbative Bogoliubov-like stochastic method described in Ref.~\cite{Deuar:13} that explicitly forbids Bose enhancement. These give $\Gamma_N$ below unity for all $N$ and all times, proving that the anisotropy $\Gamma_N>1$ seen in Fig.~\ref{fig:Ndiffbog} requires Bose stimulated scattering.

\begin{figure}
\includegraphics[clip,width=0.8\columnwidth]{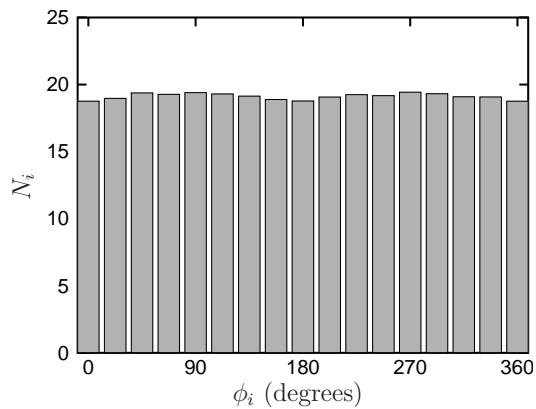}
\caption{The number of scattered atoms $N_i$ ($i=1,2,...16$) per azimuthal sector centered at $\phi_i$ at the earlier time $t=64$ $\mu$s, from the stochastic Bogoliubov simulation. Other details as in Fig.~\ref{fig:n1e5bog}. 
}
\label{fig:Nphiearly} 
\end{figure}

Another important feature in Fig.~\ref{fig:Ndiffbog} is the transient \textit{reverse}  anisotropy ($\Gamma_N<1$) that occurs for the small-$N$ cloud at early times, i.e. the number of particles at $\phi=0^{\circ}$ and $180^{\circ}$ is smaller.  
This is indicative of the trajectory deflections analyzed in Sec.~\ref{CLASS}, which initially win over Bose enhancement because the latter only becomes appreciable once the halo is sufficiently occupied. 
The low-$N$ case has the longest latency period before Bose-enhanced scattering becomes appreciable, which explains why the reverse anisotropy is seen there.  Further evidence of deflections is seen in Fig.~\ref{fig:Nphiearly} where the angular distribution of the number of scattered atoms $N_i$ versus $\phi_i$ is shown at $t=64$ $\mu$s, the time when the reverse anisotropy is strongest: 
the dip in the particle number occurs only in a narrow range around  $0^{\circ}$ and $180^{\circ}$. This is similar to what was seen in Fig.~\ref{fig:classical}, and occurs because a very elongated condensate deflects only in a narrow range of $\phi$ around its long axis. 

\subsection{Anisotropy timescales}
It is possible to make some crude estimates of the relevant timescales and from these describe the physical regimes in which parametric amplification or trajectory deflections are dominant. 
We consider a Thomas-Fermi approximation similarly to the very simple model of Sec.~\ref{PDC}, with initial chemical potential $\mu(0)=g\rho_0(0)$, and cloud radius in the short direction, $R_{\perp}(0)=\sqrt{2\mu(0)/m\omega_{\perp}^2}$. 

\begin{itemize}
\item 
The time for the two halves of the source condensate to geometrically separate in space is $t_{\rm coll}=mR_{\perp}(0)/2\hbar k_0$ 
if we ignore condensate expansion. For our system, this is approximately equal to $40$ $\mu$s.
Incorporating the spatial expansion (see the next item), which is the fastest in the transverse (or radial) direction, and monitoring the spatial separation and the disappearance of the overlap region numerically as in Sec.~\ref{CLASS} and the Appendix, gives $t_{\rm coll}\approx70$ $\mu$s.

\item The time needed for the source condensate to appreciably dilute due to transverse expansion ($t_{\rm dilution}$) can be estimated from the self-similar solution for a condensate in the Thomas-Fermi regime 
\cite{Castin-Dum}. Since the cloud width in the transverse direction grows as $\sqrt{1+(\omega_{\perp} t)^2}$, the central density dilutes by a factor of 2 at $t_{\rm dilution}=1/\omega_{\perp}\approx 140$ $\mu$s. This dilution affects the gain $\mc{G}=g\rho_0(0)/\hbar$ discussed in the simple model of Sec.~\ref{PDC}. Thus we expect $t_{\rm dilution}$ to also be the time that scattering and gain ceases, unless it has been cut off earlier by $t_{\rm coll}<t_{\rm{dilution}}$ which is the case in our system.
\item The trajectory deflections have the same physical underpinning as the transverse expansion (mean-field repulsion), and take place on the same timescale, so that $t_{\rm deflection}\approx t_{\rm dilution}$. 
\item Generally speaking, the time over which gain and hence parametric amplification has a chance to work is then $t_{\rm gain}\approx {\rm min}[t_{\rm coll},t_{\rm dilution}]$. This, however, depends also on the direction of propagation of the scattered atoms and is given in our system by the time required for the scattered atoms to escape the overlap region of the colliding condensates. Because of the anisotropy of the source condensate, the escape times are different in the $x$ and $y$-directions, and can be approximated by $t_{\mathrm{gain},x}\approx 70$ $\mu$s and $t_{\mathrm{gain},y}\approx 40$ $\mu$s as is done in the analysis of Sec. \ref{PDC}. In reality, these timescales are further reduced, which is  due to the fact that the effective gain coefficient $\cal{G}$ does not stay constant, but becomes smaller due to the dilution of the source condensates during expansion.
\item Finally, we note that the bulk of the anisotropy due to parametric amplification appears after the atoms with the short (transverse) gain-path have left the cloud, i.e. for $t\gtrsim t_{\mathrm{gain},y}\approx 40$ $\mu$s. 
\end{itemize}
Collecting this information together we can expect the following behaviour to emerge in the limiting cases. 
To achieve strong anisotropy of $\Gamma_N\gg 1$ in our current geometry, one could try to enter the regime of a larger gain and hence an exponential regime of parametric amplification of the scattering modes, governed by $n\propto \sinh^2({\cal{G}} t_{\mathrm{gain},i}/2)$ ($i=x,y$). This can, in principle, be achieved with, e.g., $\sim 5$ times larger value of the coupling $\cal{G}$, however, for the same trap frequencies, such a large value of $\cal{G}$ would require $\sim 5$ times larger peak density $\rho_0(0)$ of the source condensate. The respective initial total atom number in this case would have to be $N \sim 5\times 10^6$ (due to the Thomas-Fermi scaling of $N\propto \rho_0(0)^{5/2}$), which is about $50$ times larger than in the experiment.

Changing the trap frequencies to make the source condensate more anisotropic (e.g., by reducing only $\omega_x$ as to not influence any other relevant timescales, such as $t_{\mathrm{coll}}$) will have a much smaller overall effect as the effective escape (gain) timescales are determined not directly by the aspect ratio of the source cloud (which we note is already quite high, $\omega_{\perp}/\omega_{x}=1150/47\simeq 24.5$), but by the shape and the dynamics of the overlap region between the two condensates $t_{\mathrm{gain},x(y)}$.

To be predominately in the opposite regime of reverse anisotropy ($\Gamma_N\ll1$) one would need to reduce the collision duration to the regime of low-gain, spontaneous scattering with $t_{\rm dilution}\gg t_{\rm coll}$. Indeed, the reverse anisotropy is a result of trajectory deflections and these do not simply cease at the end of collision because they do not require atom pair creation. Therefore, if $t_{\rm dilution}\gg t_{\rm coll}$ so that exponential gain does not set in, then a situation in which primarily reverse anisotropy takes place is possible. This can be enhanced by either higher speed or lower atom number collisions. The crude analysis presented here does not let us specify quantitative values, but inspection of the numerical results of Fig.~\ref{fig:Ndiffbog} suggests that this occurs for $N\lesssim 10^4$ for our parameters.

\subsection{Atom number modulation}
\label{asym-contrib}

The cause of the much stronger angular modulation of scattered particle numbers $N_i$ in the experiment (Fig.~\ref{fig:AtomNbMod}, which gives $\Gamma_N\approx1.4$) than in the calculations is not fully understood at present. 

One possibility that we considered initially and have subsequently ruled out is a disproportionate anisotropic contribution from large-$N$ clouds. The scattering rate is proportional to density squared, so a disproportionately stronger halo is expected in denser clouds, i.e. those with larger particle number $N$. Since the number of atoms in a single run can vary by a factor of up to 3--4 between high and low $N$ values \cite{Jaskula:thesis}, the measured anisotropy of $\Gamma_N=1.4$ might be primarily due to clouds with above average $N$. 
We evaluated the anisotropy $\Gamma_N$ from sinusoidal fits to $N_i(\phi_i)$ using only restricted sets of experimental data with outlying high or low $N$ values. To sort the experimental runs we used the number of detected atoms in the halo, $N_{\rm scat}$, which is a monotonic function of $N$. Outlying low-$N$ realizations identified by $N_{\rm scat}<260$ gave an anisotropy of $\Gamma_N\approx1.3$, while outlying high-$N$ realizations with $N_{\rm scat}>360$ gave  $\Gamma_N\approx1.5$. 
We conclude that, while present, the effect is too small by itself to reconcile the difference in anisotropy between our $N=10^5$ calculation and experiment, 
considering that even the $N=5.5\times10^5$ simulation still gives an anisotropy  significantly smaller than 1.4.

All this notwithstanding, our analysis reveals the details of the competition that occurs between the gain anisotropy caused by parametric amplification, and the reverse-anisotropy caused by deflection of particles on the background mean-field of the condensate. In particular, while the reverse anisotropy occurs at short times and for small clouds, it can eventually be overcome by the gain anisotropy, and in fact usually is for the parameters of our experiment. 


\section{Relation to superradiance}
\label{SRAD}
The Bose stimulated scattering of atoms in preferred directions resembles superradiance experiments, where a single elongated atomic cloud was illuminated by light \cite{Inouye:99,Schneble:03,Moore:99,Vardi:02,Lopes:14}.
However, our purely atomic system differs from them in several important ways.

\begin{figure}
\includegraphics[width=0.52\columnwidth,clip,trim=0cm 0cm 0cm 0cm]{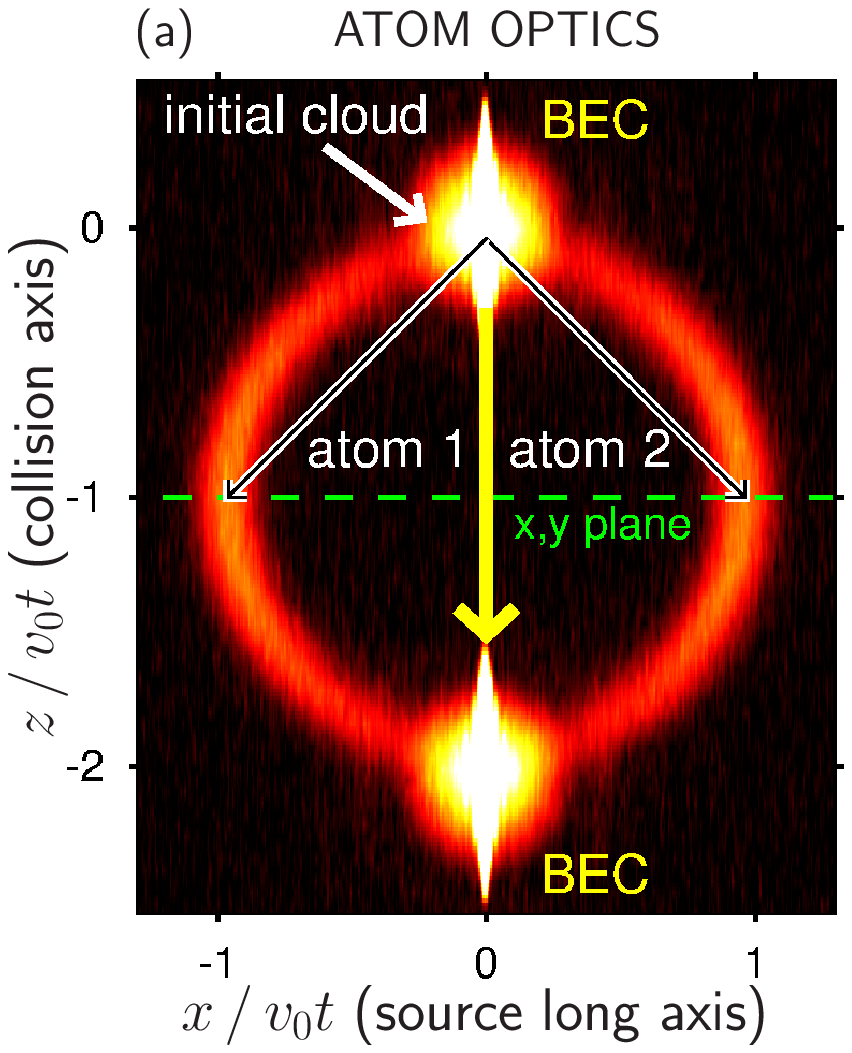}
\raisebox{0.7cm}{\includegraphics[width=0.417\columnwidth]{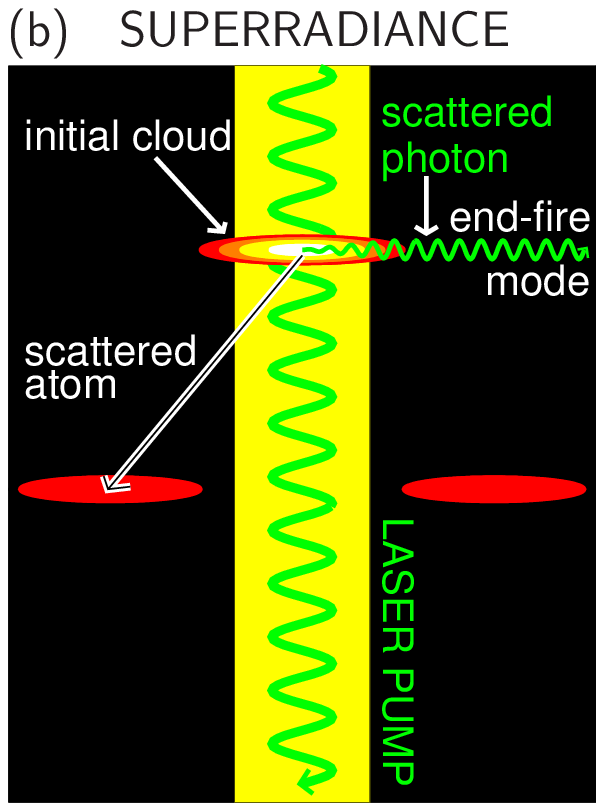}}
\caption{(Color online) 
A schematic comparison of the scattering geometry in atom collisions (a) and superradiance (b), in the rest frame of the upper condensate. The color background shows the long-time atom distribution on the $x-z$ plane. In panel (a), it comes from the the stochastic Bogoliubov simulation for the $N=10^5$ case. We use the k-space distribution at the end of the collision, scaled out to the far field via $x=(\hbar k/m) t$. The $x-y$ plane analyzed in the text is shown as a dashed green line. In both panels, arrows show flight paths.
\label{fig:srad} }
\end{figure}

For illustrative purposes, consider the schematic drawings of Fig.~\ref{fig:srad}, which compare the geometries in the two cases. 
In a superradiance setup, one has a single elongated cloud that is illuminated by a broad ``pump'' laser beam from a direction perpendicular to the long axis, which produces photon-atom scattering. 
The photons scattered along the long condensate axis preferentially stimulate more scattering of their own kind because their flight time through the condensate is longer than for photons scattered in other directions, and more gain can build up. By the time they leave the long end of the condensate they have gathered quite a few similar photons and 
strongly amplified ``end-fire'' modes get emitted along the long axes of the condensate.
Energy-momentum conservation requires that the recoiling atoms fly off at $45^{\circ}$ to the long axis.

For the purely atomic collision, the energy-momentum conservation conditions are different. 
In particular, while superradiant photons could move the whole length of their source condensates stimulating more scattering, atoms scattered in the condensate collision leave the gain (condensate overlap) region much sooner than the time it takes them to travel a distance equal to the half-length of the condensate. 

The geometry of the collision in the rest frame of the upper condensate, located at the origin, is shown in Fig.~\ref{fig:srad}(a). None of the atoms scattered into the main ring halo lie on the $x$ axis that passes through the upper condensate. This means that the paths of all the atoms that separate appreciably from the source clouds over time are inclined at a significant angle to the long axis of the condensate. Their path lengths through the condensate are accordingly reduced in comparison with the condensate half-length. 
For the $x-y$ plane that we have been analyzing in condensate collisions, {\it both} atoms are scattered at approximately $45^{\circ}$ to the collision axis ($z$), and fly at no closer angle to the long part of the condensate than $45^{\circ}$.  These closest flying atoms are shown as arrows in Fig~\ref{fig:srad}(a) and appear at $\phi=0^{\circ}$ and $\phi=180^{\circ}$ in Figs.~\ref{fig:DensityData} -- \ref{fig:n1e5bog}.
They remain in the condensate $\sim\sqrt{2}$ times longer than atoms scattered along the perpendicular, into-the-plane $y$ direction, which end up at $\phi=90^{\circ}$ and $\phi=270^{\circ}$. This factor is responsible for the gain anisotropy in the parametric amplification process, as explained in Sec.~\ref{PDC}.
Related restrictions to scattering angles for atom pairs have also been reported for molecular dissociation setups \cite{Vardi:02}.

Another difference is that with the two clouds acting as the coherent pumps for each other, the scattering lasts only as long as the duration of the collision, which is limited both by the narrow transverse width of the cloud in the $z$ direction, and by the loss of density from transverse expansion. For a strongly elongated cloud, this collision timescale is much shorter than the time for any scattering products to travel the length of the condensate, regardless of scattering angle restrictions.

Both of these effects (reduced illumination time and inability to scatter along the long axis of the condensate) contribute to a reduction of the anisotropy of the gain in a purely atomic collision compared to the superradiance stimulated by a light source.


\section{Conclusions and outlook}
\label{CONC}

Unlike the anisotropy described in the work of Krachmalicoff {\it et al. }\cite{Krachmalnicoff:10} that was best understood as a particle effect, the anisotropy in the same experiment that we have discussed here 
does have a qualitative analogue in optical phenomena. BEC collisions, then, produce anisotropies of both particle-like and wave-like nature. The effect described in the present paper bears a number of hallmarks of optically pumped superradiance (such as stimulated scattering enhanced along the long direction of the condensate cloud, and pair correlations among the scattering products), however we have also shown that the similarity holds only to a limited degree. 

We have found the existence of a competition between the anisotropy in the gain due to parametric amplification by the atom clouds, and the reverse-anisotropy caused by deflection of particles on the background mean-field of the condensates. In particular, while the reverse anisotropy occurs at short times and for small clouds, it is usually eventually overcome by the gain anisotropy. Furthermore, the time over which gain occurs and the energy-momentum conservation conditions in an atom cloud collision are markedly different from those in superradiance, which makes it relatively much more  difficult to achieve very strong end-fire modes with elongated condensates.

Different geometries however, can lead to stronger gain-induced anisotropies. Consider the collision of two oblate,
``pancake-shaped'' condensates, which collide along a radial axis: suppose, for example, that the pancakes are flattened along
the $x$ direction and collide along the $z$ direction. Then, in the rest frame of one of the condensates, the $45^{\circ}$ cone around
the collision axis ($z$) allowed by energy-momentum conservation still intersects a long dimension of the condensates on the $y-z$ plane. Atoms scattered in this direction still have a long flight path through the condensate. They end up near  $\phi=90^{\circ}, 270^{\circ}$ on the $k_x-k_y$ plane. With an aspect ratio of $\lambda\gg1$ for the condensate, they have a flight path through the condensate that is $\lambda/\sqrt{2}$ times longer than those scattered towards $\phi=0^{\circ}, 180^{\circ}$.  The gain then leads to strongly Bose enhanced scattering near $\phi=90^{\circ}$ and $270^{\circ}$.

The results presented in this work provide the initial answers to a question that has been posed in the atom optics field for over a decade ---  whether or not spontaneous directionality is achievable in the case of atom-atom pair emission from an elongated atom cloud \cite{Vardi:02,Ogren-directionality}. Its answer is important for both fundamental and applied considerations. For example, atom pairs scattered into vacuum have different and more strongly nonclassical properties than those whose scattering has been seeded in a  four-wave mixing process such as reported in \cite{Deng:99}. It is advantageous to have such pairs collimated in space. Our results demonstrate that such spontaneous directionality is achievable, but the conditions are appreciably different than in optically pumped superradiance.

Finally, in a broader context, one expects other situations that generate a scattered atom halo, such as molecular dissociation in a condensate 
\cite{Poulsen:01,Durr:04,Greiner:05,Savage:06,Savage:07,Ogren:08,Davis:08,Ogren-directionality,Ogren:10}, 
atomic parametric down-conversion 
\cite{Bucker:11,RuGway:11,Bonneau:13,Dall:09,Campbell:06,Gemelke:05},  
or the interaction of a condensate with barriers and obstacles 
\cite{Pasquini:06,Carusotto:06,Sykes:09,Scott:06,Scott:07}, 
to also be susceptible to the same anisotropy-producing processes.

\begin{acknowledgments}
We acknowledge discussions with  P. B. Blakie, M. Trippenbach, P. Zi\'{n}, T. Wasak, and J. Chwede\'nczuk. 
This work was supported by the ANR ProQuP project, the  IFRAF-nano-K program, the Triangle de
la Physique and the Labex PALM. P.D. acknowledges the support of the National Science Centre grant No. 2012/07/E/ST2/01389, K.V.K acknowledges  support by the ARC Future Fellowship grant No. FT100100285. 
\end{acknowledgments}


\appendix*
\section{A classical test-particle treatment}
\label{SKIIER}

In the classical test-particle method, the binary collisions that
produce the $s$-wave scattering halo are mimicked by random events
that create pairs of particles of mass $m$ with equal but opposite
momenta $\mathbf{k}$ and $-\mathbf{k}$ in the center-of-mass frame of the BECs. The initial direction of flight is generated randomly, and is 
distributed isotropically with respect to $\phi$, reflecting the isotropic nature of $s$-wave collisions. 
The absolute values of the momenta, on the other hand, are drawn randomly from a Gaussian probability
distribution centered at $|\mathbf{k}|=k_{0}$ and having a
width that varies with the polar angle: the annular variation of the
width respects the quantum mechanical momentum uncertainty of the
source condensate in different directions.

The initial positions of the pairs are also generated randomly, weighted 
using a probability distribution
that is proportional to $\rho_0(\mathbf{x})^{2}$, where $\rho_0(\mathbf{x})$
is the density profile of the initial $t=0$ condensate. 
This takes into
account the fact that the probability of scattering and hence the
probability of pair creation is proportional to the product of the densities of the split condensates $\rho_{1}(\mathbf{x},t)\rho_{2}(\mathbf{x},t)$,
which at time $t=0$ is proportional to $\rho_0(\mathbf{x})^{2}/4$.
The density $\rho_1(\mathbf{x},t)$ itself, is approximated by an inverted parabola as in the Thomas-Fermi approximation.

Each test particle is then propagated in time according to the classical
Newtonian equations, subject to the effective external potential formed
by the mean-field of the colliding condensates. We monitor the classical
trajectories of the test particles and record their final momentum
distribution after their escape from the collision zone into far regions without a mean-field
potential.

\subsection{Static potential}

\begin{figure}
\includegraphics[width=0.6\columnwidth,clip]{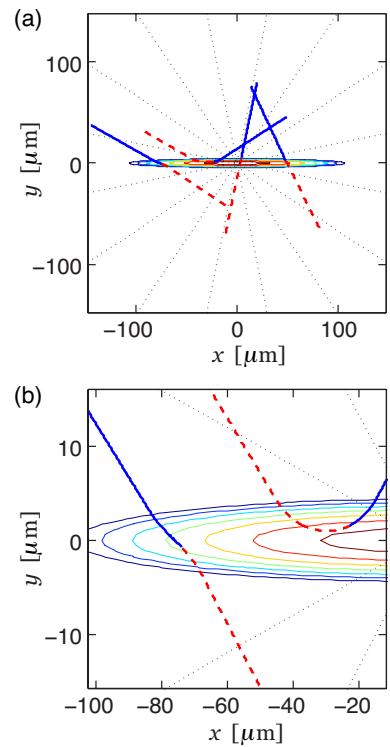} 
\caption{(Color online) Examples of simulated classical test particle trajectories. Solid blue and dashed red lines show the paths followed by particles in a pair for 1 ms. 
The pair is produced at the point where the two lines meet within the
collision zone that is shown as a contour plot. Subplot (b) magnifies detail near the source zone. }
\label{fig:trajectories} 
\end{figure}

To illustrate the effect of the mean-field potential in its simplest
form, we first consider a \textit{static} potential given by $U(\mathbf{x})=2g\rho_0(\mathbf{x})$,
where $g=4\pi\hbar^{2}a/m$ is the coupling constant for the binary
$s$-wave interactions, and $a$ is the $s$-wave scattering length.
This form of the mean-field potential follows from the Bogoliubov
analysis of the collision problem (see Eq.~(\ref{Bog})\,), which
in the classical limit corresponds to the motion of a test particle
in an external potential $U(\mathbf{x})=2g\rho_0(\mathbf{x})$.
The problem is analytically solvable for the parabolic shape of $U(\mathbf{x})=2g\rho_0(\mathbf{x})$
in the Thomas-Fermi approximation:  
\begin{equation}
\rho_0(\mathbf{x})= \text{max}\left[\ 
\rho_0(0)\left(1-\frac{x^{2}}{R_{x}^{2}}-\frac{y^{2}}{R_{y}^{2}}-\frac{z^{2}}{R_{z}^{2}}\right), \quad 0\ \right].
\end{equation}
with $R_{i}$ ($i=x,y,z$) being the Thomas-Fermi radii
and $\rho_0(0)$ the peak density.

For simplicity, we only consider a 2D dynamics corresponding to monitoring
the test particles with a strictly zero $z$-component of the initial
momentum. The classical trajectories of such particles remain in the
$x-y$ plane. In Fig.~\ref{fig:trajectories} we show examples
of such trajectories after $1$ ms of rolling down the mean-field
potential $U(\mathbf{x})$. Typical deflections away from the axial
($x$) direction due to the mean-field potential are shown. The strongest effect is from the transverse gradient. The most dramatic deflection scenario
corresponds to the case of a complete reflection from the potential
hill, as illustrated in one of the examples of Fig.~\ref{fig:trajectories}(a)
[with Fig.~\ref{fig:trajectories}(b) showing the magnified version].

\begin{figure}
\includegraphics[width=0.85\columnwidth,clip]{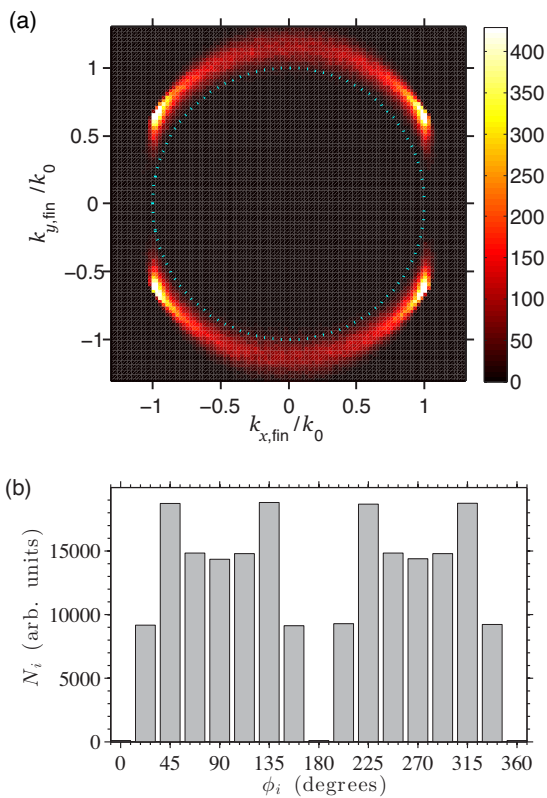} 
\caption{(Color online) 
Halo properties in a classical test-particle model as in Fig. \ref{fig:classical}, but for a static mean-field potential.
}
\label{fig:classical-static} 
\end{figure}

In order to generate statistically significant data we typically evolve
about $10^{6}$ test particles. The resulting distribution of final
momenta is shown in Fig. \ref{fig:classical-static}(a) where we
see vanishingly small densities along the $x$ axis (i.e., at polar
angles of $0^{\circ}$ and $180^{\circ}$), together with four high-density,
``focal'' regions away from the $x$ axis. This peculiar shape originates
from trajectory deflections away from the axial direction of the potential
$U(\mathbf{x})$. To further quantify the data, we bin the resulting
momentum distribution into $16$ angular bins and plot the resulting
total atom number $N_i$ ($i=1,2,...16$) in each bin as a
function of the polar angle $\phi_{i}$. The effect of trajectory
deflections manifests itself as a nontrivial deep modulation of $N_{i}$,
with minima at $\phi_{i}=0^{\circ}$ and $180^{\circ}$ ($\Gamma_N\ll 1$), and is shown in Fig.~\ref{fig:classical-static}(b).

The relevant physical parameters in our analysis are as in the experiment.
We consider an initial condensate of $N\sim\!\!10^{5}$ helium-$4$
atoms, in the state $m_{x}=1$ with an $s$-wave scattering length
of $a=7.51$~nm, trapped in a harmonic trap with frequencies of $\omega_{x}/2\pi=47$
Hz and $\omega_{y}/2\pi=\omega_{z}/2\pi=1150$ Hz. The condensate
initial density profile is approximated by the Thomas-Fermi inverted
parabola, with Thomas-Fermi radii $R_{x}=114$~$\mu$m and $R_{y,z}=4.67$~$\mu$m
and peak density $\rho_0(0)=2.4\times10^{19}$~m$^{-3}$. The
average initial speed of the test particles is $v_{0}=7.31$~cm/s (momentum $k_0=mv_0/\hbar=4.61\times 10^6$ m$^{-1}$).
With these parameters, the initial average kinetic energy per particle,
$E_{\mathrm{kin}}^{\mathrm{(ini)}}=\hbar^{2}k_{0}^{2}/2m$, is $2.35$
times larger than the peak of the potential $U(0)$. For a test particle
starting to roll down from the very top of the potential hill with
momentum $k_{0}$, the final kinetic energy $E_{\mathrm{kin}}^{\mathrm{(fin)}}=\hbar^{2}k_{\mathrm{fin}}^{2}/2$
is equal to $E_{\mathrm{kin}}^{\mathrm{(fin)}}=E_{\mathrm{kin}}^{\mathrm{(ini)}}+U(0)$,
implying that the maximum final speed of the test particles is $k_{\mathrm{fin}}^{(\mathrm{max})}\simeq1.2k_{0}$.
For test particles created on the side of the potential hill 
the final momentum will be between $k_0$ and $k_{\mathrm{fin}}^{(\mathrm{max})}$, as seen in the average radius and width of the halo in Fig.~\ref{fig:classical-static}(a).

\subsection{Time-dependent potential}

The full classical particle calculation whose results are shown in Fig.~\ref{fig:classical}, took into account a \textit{time-dependent} mean-field potential $U(\mathbf{x},t)$. 
This case is no longer analytically solvable, and we generate the
numerical data by solving the Newtonian equations of motion using
the velocity Verlet algorithm \cite{Swope:82} (for a recent use of
the method in the context of ultracold atoms, see \cite{Lepers:10}).
The time-dependent potential $U(\mathbf{x},t)$ at each time step
is closely approximated by the actual overlap region---on the $x-y$
plane---between the colliding and spatially separating condensates.
For numerical simplicity, we approximate the overlap region by an
inverted parabola at each time step; the overlap dwindles with time and eventually disappears on a time scale of $70$~$\mu$s. 
This time scale is approximately the actual duration required for
the colliding condensates to geometrically separate for our choice
of parameters.
Note that in our approximation, the interference fringes between the two counter-propagating
condensates have been averaged out. This is allowable because the characteristic size of the quantum
mechanical wavefunction of the scattered atoms in a binary $s$-wave
collision is of the same order as the size of the source condensate, which in turn is much greater than the fringe spacing. Therefore, the scattered atoms see an averaged mean-field potential.
 As previously, we  stop the simulation at $t_{\max}=1$~ms
and record the final momenta of all test particles.

The results for the final momentum distribution and the binned atom
number $N_i$ as a function of the polar angle $\phi_i$ are shown in Fig.~\ref{fig:classical}.
The trajectory deflections are now a weaker effect than
with a static potential, in that the variations in $N_i$ are much less extreme, and the final momentum distribution
is qualitatively much closer to the one observed experimentally.
The annular variation of the width and the peak
density is largely a reflection of the initial momentum distribution
of the test particles, which has this same kind of anisotropy in
the width, but scatters the same flux in all directions. The peak height in the halo then compensates for the width variation. 

Despite the greater similarity with the experimentally observed
momentum distribution (cf. Fig.~\ref{fig:DensityData}), the final
distribution in Fig.~\ref{fig:classical} still contains the effects
of classical trajectory deflections \textit{away} from the $x$ axis ($\Gamma_N<1$ -- the reverse anisotropy).
One has a near-sinusoidal modulation of the binned atom
number distribution $N(\phi_{i})$ with minima at $0^{\circ}$ and $180^{\circ}$, even though
the respective initial distribution is isotropic. However, the experimentally observed modulation has maxima, rather than minima,
at these angles.

\bibliography{gain_anisotropy}

\end{document}